\documentclass[pdftex]{pasj01}
\Received{2023/07/04}%{yyyy/mm/dd}
\Accepted{2023/08/02}%{yyyy/mm/dd}
\usepackage[dvipdfmx]{}
\usepackage{enumerate}
\usepackage{color}
\usepackage{pslatex} % to use PostScript fonts
\usepackage{lscape}
\usepackage[switch,mathlines]{lineno}

\newcommand{\twelveco}{\mbox{$^{12}$CO}} 
\newcommand{\thirteenco}{\mbox{$^{13}$CO}} 
 
\newcommand{\twelvecol}{\mbox{$^{12}$CO($J$=1--0)}} 
\newcommand{\thirteencol}{\mbox{$^{13}$CO($J$=1--0)}} 
 
\newcommand{\thirteencoh}{\mbox{$^{13}$CO($J$=2--1)}} 
\newcommand{\twelvecohh}{\mbox{$^{12}$CO($J$=3--2)}}

\newcommand {\msun}{\mbox{M$_\odot$}}

\newcommand {\kms}{\mbox{km~s$^{-1}$}}
\newcommand {\kkms}{\mbox{K~km~s$^{-1}$}}
\newcommand {\vlsr}{\mbox{$V_\mathrm{LSR}$}}
\newcommand {\nhtwo}{\mbox{$N_\mathrm{H_2}$}}

\newcommand {\vcen}{\mbox{$V_\mathrm{center}$}}

\newcommand {\dv}{\mbox{$dV$}}

\newcommand {\htwo}{\mbox{H{\sc ii}}}

\newcommand {\riso}{\mbox{R$_{13/12}$}}
\newcommand {\vrep}{\mbox{$V_\mathrm{rep}$}}

\begin{document}

\title{Discovery of a molecular cloud possibly associated with the youngest Galactic SNR G1.9$+$0.3}

\author{Rei \textsc{Enokiya}\altaffilmark{1,2,3}%
\thanks{Example: Present Address is xxxxxxxxxx}}
\altaffiltext{1}{Department of Physics, Faculty of Science and Technology, Keio University, 3-14-1 Hiyoshi, Kohoku-ku, Yokohama, Kanagawa 223-8522, Japan}
\altaffiltext{2}{National Astronomical Observatory of Japan, Mitaka, Tokyo 181-8588, Japan}
\altaffiltext{3}{Faculty of Engineering, Gifu University, 1-1 Yanagido, Gifu, Gifu 501-1193, Japan}
\email{enokiya@keio.jp, enokiya.rei.t4@f.gifu-u.ac.jp}

\author{Hidetoshi \textsc{Sano}\altaffilmark{2,3}}

\author{Miroslav, D. \textsc{Filipovi{\'c}}\altaffilmark{4}}
\altaffiltext{4}{Western Sydney University, Locked Bag 1797, Penrith South DC, NSW 2751, Australia}

\author{Rami, Z. E. \textsc{Alsaberi}\altaffilmark{4}}

\author{Tsuyoshi \textsc{Inoue}\altaffilmark{5}}
\altaffiltext{5}{Department of Physics, Konan University, Okamoto 8-9-1, Higashinada-ku, Kobe, Hyogo 658-8501, Japan}

\author{Tomoharu \textsc{Oka}\altaffilmark{1}}


\maketitle

%============================Abstract (below)==============================
\begin{abstract}
The youngest known Galactic supernova remnant (SNR) G1.9$+$0.3 has high-velocity supernova shock beyond 10000~$\kms$, and it is considered to be one of the major candidates of a PeVatron.
Despite these outstanding properties, the surrounding interstellar matter of this object is poorly understood.
We investigated the interstellar gas toward G1.9$+$0.3 using the $\twelvecohh$ data with the angular resolution of 15$\arcsec$ obtained by the CHIMPS2 survey by the James Clerk Maxwell Telescope, and discovered three individual clouds at $-$1, 7, and 45~$\kms$.
From its morphological and velocity structures, the $-$1~$\kms$ cloud, having the largest velocity width $>$20~$\kms$ and located at the distance of the Galactic Center, is possibly associated with the SNR.
The associated cloud shows a cavity structure both in space and velocity and coincides well with the SNR.
We found that the associated cloud has higher column densities toward three bright, radio synchrotron-emitted rims where the radial expansion velocity of the supernova shock is decelerated, and the cloud is faint in the other parts of the SNR.
This is the first direct evidence indicating that the highly anisotropic expansion of G1.9$+$0.3 observed by previous studies results from the deceleration by the interaction between the supernova shock and surrounding dense interstellar medium.
\end{abstract}
%============================Abstract (above)==============================
%\pagewiselinenumbers

%============================introduction (below)==============================
\section{Introduction} \label{sec:intro}
%%%%%%%%%%%%%%%%%%%%%%%%%%%%%%%%%%%%%%%%%%%%%%%%%%%%%%%%%%%%%%%%%
\begin{figure*}[t]%Fig1	X-radio
\begin{center}
\includegraphics[width=10cm]{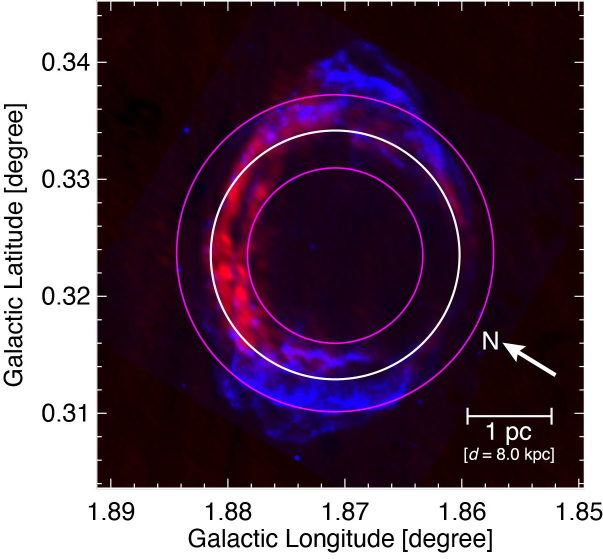}
\end{center}
\caption{Two-color composite image of G1.9$+$0.3, where red is 9~GHz radio continuum emission and blue is 2--7 keV emission. The white arrow indicates the direction of the north. The white and magenta circles indicate the SN-shell radius (=38$\farcs$5, see Figure 5 in \cite{luk20}) and the outer/inner boundary of the SN shell, respectively, from the dynamical center ($l$, $b$) = (1$\fdg$8710, 0$\fdg$3237) \citep{bor17}. The width from the outer (inner) boundary to the shell center (=11$\farcs$3) was defined as the half-width at half-maximum of the radio shell fitted by a Gaussian function \citep{luk20}.}\label{g1.9}
\end{figure*}
%%%%%%%%%%%%%%%%%%%%%%%%%%%%%%%%%%%%%%%%%%%%%%%%%%%%%%%%%%%%%%%%%

Supernova remnants (SNRs) are believed to be cosmic-ray accelerators in the Universe (e.g., \cite{fil21,row21}).
Since cosmic rays are accelerated beyond PeV only at the early stage of an SNR evolution with the shock velocity $>$10,000~$\kms$ \citep{hes14}, investigations of young SNRs are very important for astrophysics.

G1.9$+$0.3 (hereafter G1.9; \cite{gre84}) is the youngest known Galactic SNR toward the Galactic Center (GC) with an age of $\sim$100 year \citep{rey08,bor17,deh14,luk20}.
This SNR is believed to be located in the Galactic Center region (e.g., \cite{car11}).
Although TeV and GeV gamma-ray emissions have not been detected toward G1.9 so far by the current generation of telescopes, possibly owing to its large distance and young age, it is expected to provide effective particle acceleration due to its highest expanding velocity among the known Galactic SNRs.
Thus, G1.9 attracts keen interest as the prime candidate for the Galactic cosmic-ray accelerator to PeV energies (i.e., PeVatron; \cite{aha17}).

\citet{bor17} observed a lot of X-ray filaments, which constitute X-ray shell of the SNR and likely trace supernova (SN) shock fronts, from 2011 to 2015 and measured the distribution of the radial expansion velocity (see Figure~3 in \cite{bor17}).
Despite the young age, there is a factor of almost 5 velocity difference depending on the directions---the south-east and north-west parts of the SNR where the X-ray filaments are very bright (hereafter X-ray-bright rims) show higher expanding velocity, while the south-west, north, and north-east parts where radio emission is very bright (hereafter radio-bright rims) show slower expanding velocity---and \citet{bor17} pointed out that the radial expansion is highly anisotropic.
A similar trend has been confirmed by radio continuum observations compiled over 30 years \citep{deh14,luk20}.
Two possible origins for the anisotropic expansion of the SN shock have been proposed so far.
One is an interaction model: the anisotropic expansion has been caused by the interaction with surrounding material, and thereby is an acquired cause \citep{bor17}.
The other is a non-spherical SN explosion model: the anisotropic expansion originated from the non-spherical SN explosion, and thereby is a congenital cause \citep{gri21}.

Another mystery of G1.9 is the anti-correlation between the radio and X-ray emissions.
It may be relevant to the origin of the marked anisotropic expansion.
The SNR G1.9 in radio continuum emission is bright, bilaterally asymmetric and peaked at the south-west, north, and north-east. In X-rays, it bilaterally peaks at the north-west and south-east (Figure~\ref{g1.9}) sides that are parallel to the Galactic plane.
Considering that all SNR emission comes from synchrotron radiation (e.g., \cite{rey08}), this anti-correlation is puzzling.

Based on the distribution of the anisotropic expanding shock velocities, which shows that slow shocks are concentrated toward the radio-bright rims and rapid shocks are concentrated toward the X-ray-bright rims, \citet{bor17} proposed a deceleration scenario such that the anisotropic expansion was achieved by the encounter of the dense surrounding material.
In the dense material region, given the electron fraction in all particles is $\sim$10$^{-4}$ (e.g., \cite{ell05}), the electron density is expected to become larger, while the maximum energy becomes lower since $E_\mathrm{max} \propto B_{\nu} v^{2}_\mathrm{shock}$ for the age-limited acceleration (e.g., \cite{rey08,gri21}).
Therefore, the resulting synchrotron radiation is bright in radio compared to the rapid shock region, which is on the contrary bright in X-ray \citep{bor17}.
However, there is still no direct evidence of the dense gas interacting with the shocks at the slower regions, meaning that we cannot rule out the possibility of the anisotropic explosion scenario.
To resolve the above questions, investigations of gas distribution in the vicinity of the SNR G1.9 is essential.
We study interstellar gas toward G1.9 through molecular line emissions in this paper.

The paper is organized as follows.
Section 2 describes observations and data reductions, while section 3 presents the results of our investigation.
We discuss the possible origin of the molecular gas, radio continuum, and X-ray in section 4.
Section 5 summarizes our findings.
%============================introduction (above)==============================

%============================Datasets (below)==============================
%\section{Datasets} \label{sec:data}
\section{Data and analyses} 
 \label{obs}

 \subsection{Molecular lines}
As the main molecular gas tracer, we use $\twelvecohh$ observations obtained with the CHIMPS2 survey by the James Clerk Maxwell Telescope (JCMT) \citep{ede20}.
The angular resolution, velocity resolution, and typical r.m.s. noise fluctuation are 15$\arcsec$, 1.0~$\kms$, and 0.8 K, respectively.
The details of the observations and the data reduction are fully described in \citet{ede20}.
As the dense gas tracer, we use $\thirteencoh$ data with the angular resolution of $\sim$20$\arcsec$ obtained with the SEDIGISM survey by the Atacama Pathfinder EXperiment (APEX) telescope.
The full description of the observations is seen in \citet{sch21}.
We also use 1667~MHz OH data obtained with the Green Bank telescope (proposal ID: GBT/09B-007, PI: Natsuko Kudo). 
The angular resolution, velocity resolution, grid size, and typical r.m.s noise are 8$\arcmin$, 1~$\kms$, 4$\arcmin$, and $\sim$0.2 K, respectively.

The $\twelvecol$ and $\thirteencol$ data sets with the angular resolutions of $\sim$21$\arcsec$ obtained with the Nobeyama 45m telescope were used to study $\thirteenco$/$\twelveco$ ($\riso$) in molecular clouds toward the central molecular zone.
The coverage of these datasets are ($l$, $b$) = ($-$0\fdg8 to 1\fdg4, $-$0\fdg35 to 0\fdg35), thus do not cover the G1.9 region.
The full description of the observations is in \citet{tok19}.

\subsection{Radio continuum observations}
We analysed Australia Telescope Compact Array (ATCA; project code C1952) observations that are summarized in Table~\ref{tab:summary_obs}.
All observations were carried out in ``snap-shot'' mode, with one hour of integration over a 12-hour observing session.
The Compact Array Broadband Backend (CABB) was used with 2048\,MHz bandwidth and 2049\,channels at wavelengths of 3\,cm ($\nu$~=~8000--10000\,MHz; centered at 9000\,MHz) totaling 613.8\,min of integration. 

We used \textsc{miriad}\footnote{http://www.atnf.csiro.au/computing/software/miriad/} \citep{sau95} and \textsc{karma}\footnote{http://www.atnf.csiro.au/computing/software/karma/} \citep{goo95} software packages for reduction and analysis.
All observations were calibrated using the phase and flux calibrators listed in Table~\ref{tab:summary_obs} with two rounds of self-calibration using the \textsc {selfcal} task.
Imaging was completed using the multi-frequency synthesize \textsc{invert} task with natural Briggs weighting (robust~=~0).
The \textsc {Clean} and \textsc {Restor} algorithms were used to deconvolve the images, with primary beam correction applied using the \textsc{linmos} task.
The rms for the 9000\,MHz image is $\sim$26\,$\mu$Jy\,beam$^{-1}$ with a synthesized beam of $3\farcs1\times1\farcs0$ and PA of $-$11$\fdg$7.

%%%%%%%%%%%%%%%%%%%%%%%%%%%%%%%%%
\begin{table*}[t]
	\caption{Summary of ATCA observations for SNR G1.9+0.3.}
	\centering
	\label{tab:summary_obs}
	\begin{tabular}{@{}cclllcc@{}}
		\hline
    Observing   & Array    &  Frequency $\nu$  & Flux           & Phase              & Integrated time \\ % & Reference \\
    date        & config.  &  (MHz)        & calibrator         & calibrator        & (minutes) \\
		\hline
	 2018~October~19 & 6A        &   9000   & PKS\,B1934--638   &  PKS\,B1710--269  & 206.4  \\	
	 2018~December~28 & 1.5D      &   9000   & PKS\,B1934--638   & PKS\,B1710--269   & 147.6 \\
     2019~May~11 & 1.5B      &   9000  & PKS\,B0823--500   &  PKS\,B1710--269  & 259.8 \\    
		\hline
	\end{tabular}
\end{table*}
%%%%%%%%%%%%%%%%%%%%%%%%%%%%%%%%%

\subsection{X-ray}
We used archival X-ray data obtained by Chandra with Advanced CCD Imaging Spectrometer S-array.
We combined 26 individual observations from 2007 February (ObsID: 6708) to 2015 September (ObsID: 18354) using Chandra Interactive Analysis of Observations (CIAO; \cite{fru06}) software version 4.12 with CALDB 4.9.1 \citep{gra07}.
The archival data have been published in previous papers \citep{rey08,car11,bor13,bor14,zog15,aha17,tsu21}.

All the data were reprocessed using the ``chandra\_repro'' procedure.
We created an energy--filtered, exposure-corrected image using the ``merge\_obs'' procedure in the energy band from 0.5 to 7.0 keV.
The resulting effective exposure is $\sim$1.66 Ms in total.
%============================Datasets (above)==============================

%============================Results (below)=======================
\section{Results} 
 \label{sec:res}%3.1

\subsection{Molecular clouds toward the SNR}

%%%%%%%%%%%%%%%%%%%%%%%%%%%%%%%%%
\begin{figure*}[t]%Fig2	specta
\begin{center}
\includegraphics[width=15cm]{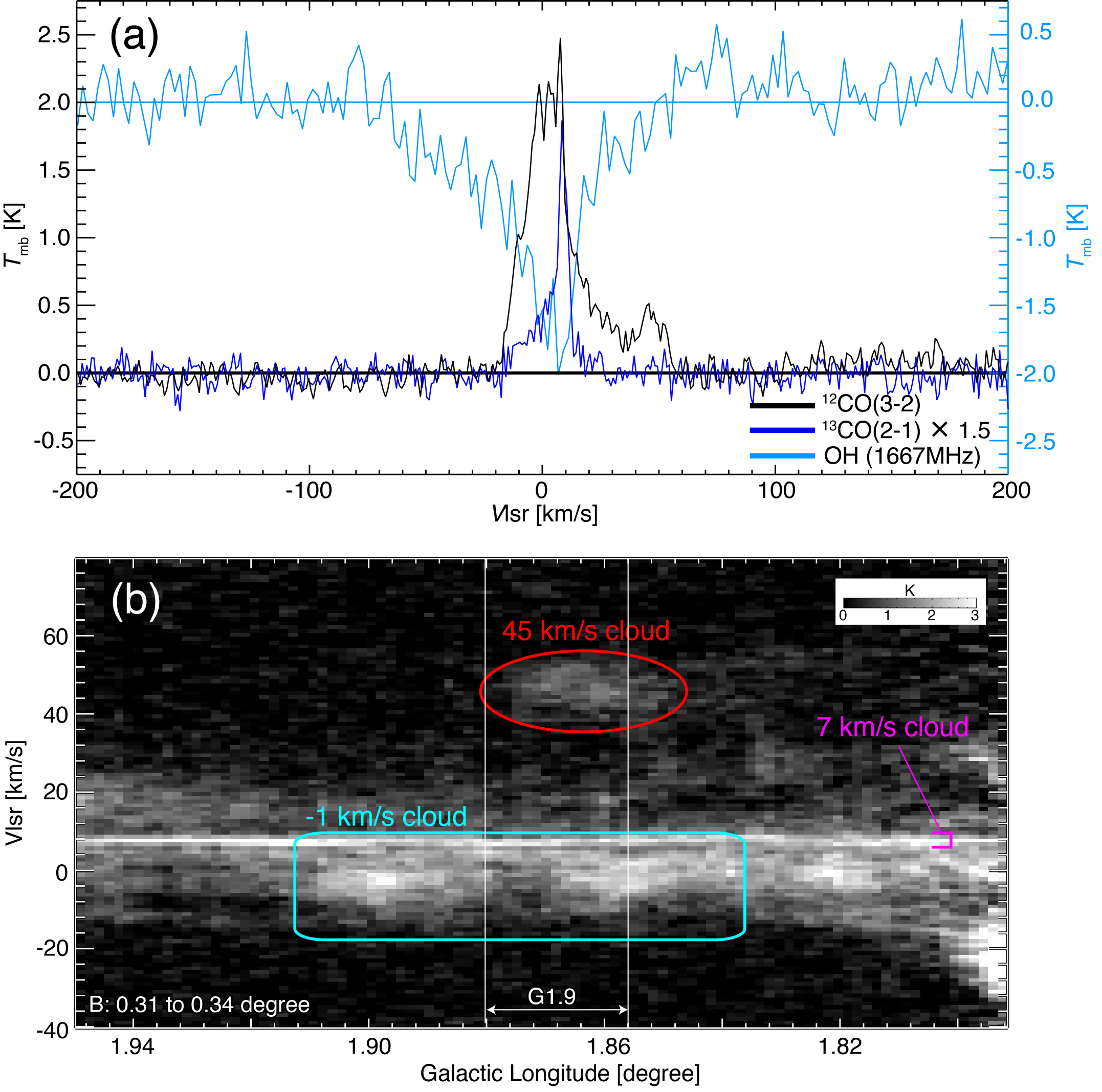}
\end{center}
\caption{(a) Averaged spectra of $\twelvecohh$ (black), $\thirteencoh$ (blue), and OH 1667~MHz (light blue) within ($l$, $b$) = (1\fdg833 to 1\fdg900, 0\fdg300 to 0\fdg367). The averaged area is indicated by the black rectangle at the top left-hand panel in figure~\ref{lbch}. (b) Longitude-velocity diagram toward G1.9. The integrated latitude range is indicated at the top left. The three molecular clouds are labelled with three different colors.}\label{spec}
\end{figure*}
%%%%%%%%%%%%%%%%%%%%%%%%%%%%%%%%%

%%%%%%%%%%%%%%%%%%%%%%%%%%%%%%%%%
\begin{figure*}[t]%Fig3	LBch_1
\begin{center}
\includegraphics[width=15cm]{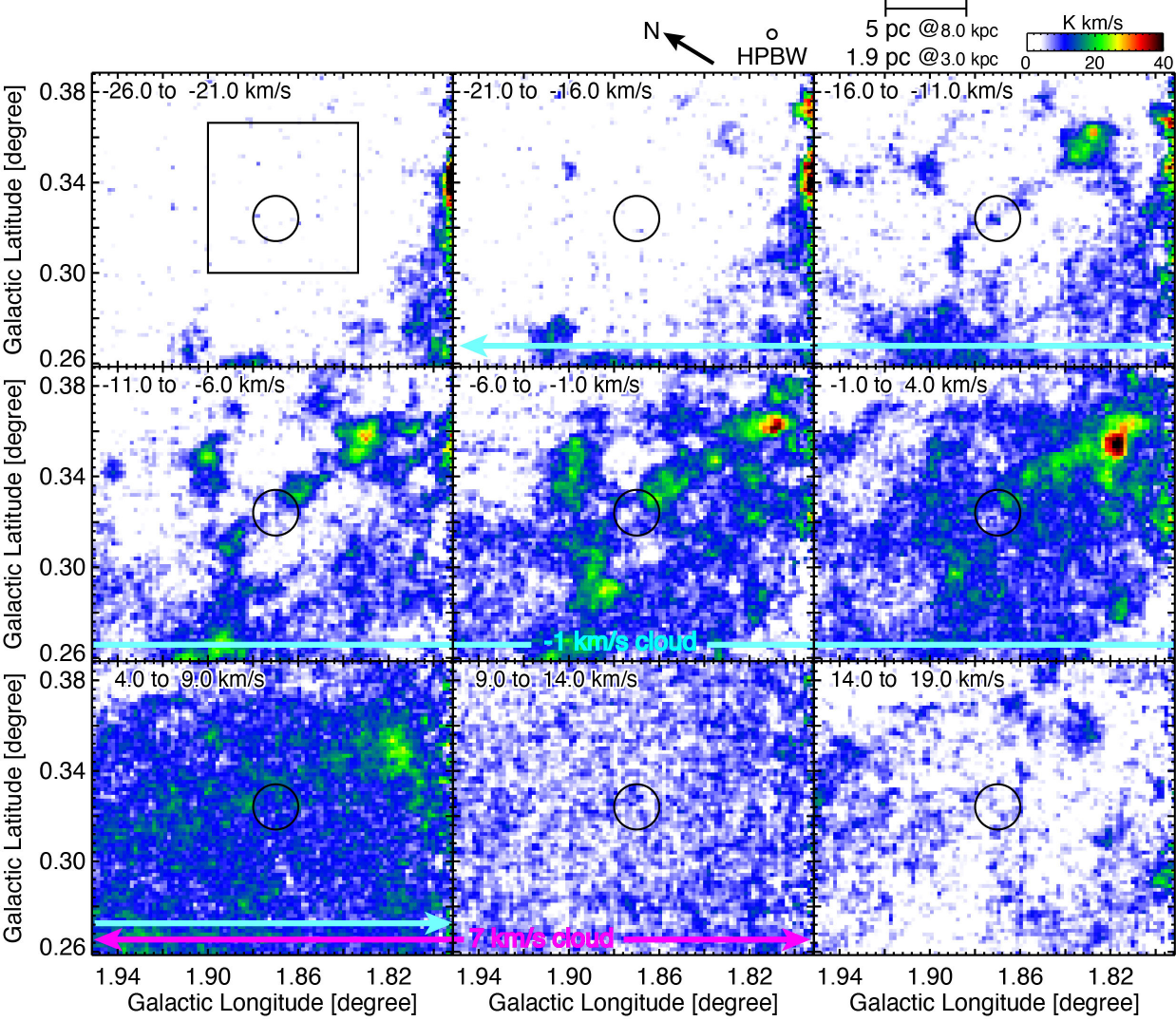}
\end{center}
\caption{Velocity channel distribution of the 1\fdg5 by 1\fdg2 area toward G1.9 in $\twelvecohh$. The black circle indicates the position of the SNR. The half-power-beam-width (HPBW), 5 pc scale bar, and the north direction are indicated at the top of Figure. The black rectangle at the top-left panel corresponds with the area used for averaged spectra in Figure~\ref{spec}. The velocity ranges for the three clouds are indicated by cyan, magenta and red arrows at the bottom of the panels. Note that the velocity channel was rebinned in advance for Figure~\ref{lbch} to show the three molecular clouds, and the arrows are guides only and do not reflect exact values of velocity ranges for the clouds.}\label{lbch}
\end{figure*}
%%%%%%%%%%%%%%%%%%%%%%%%%%%%%%%%%

%%%%%%%%%%%%%%%%%%%%%%%%%%%%%%%%%
\setcounter{figure}{2}
\begin{figure*}[t]%Fig3	LBch_2
\begin{center}
\includegraphics[width=15cm]{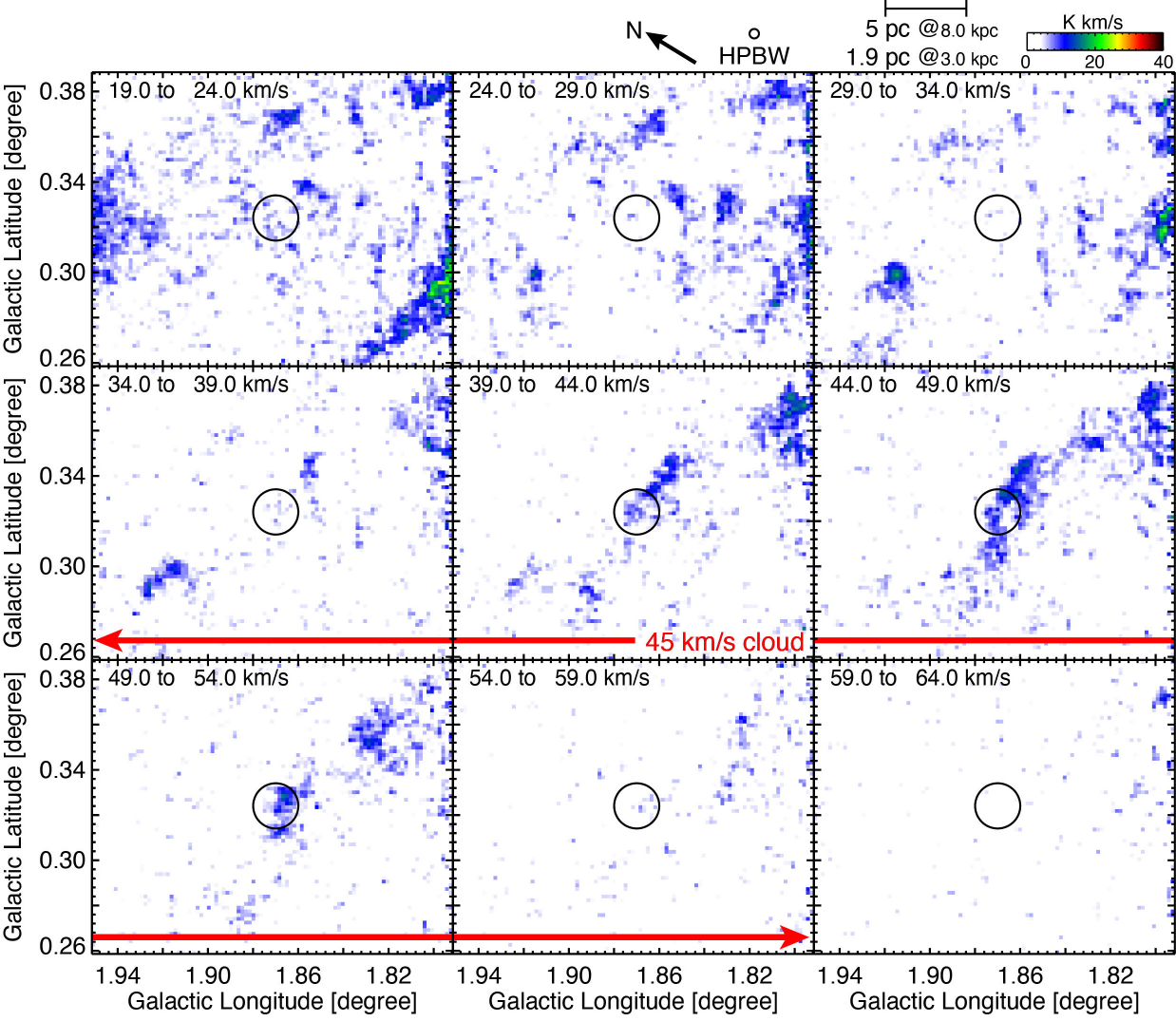}
\end{center}
\caption{$Continued$}
\end{figure*}
%%%%%%%%%%%%%%%%%%%%%%%%%%%%%%%%%

%%%%%%%%%%%%%%%%%%%%%%%%%%%%%%%%%
\begin{figure*}[t]%Fig4	clouds
\begin{center}
\includegraphics[width=\linewidth]{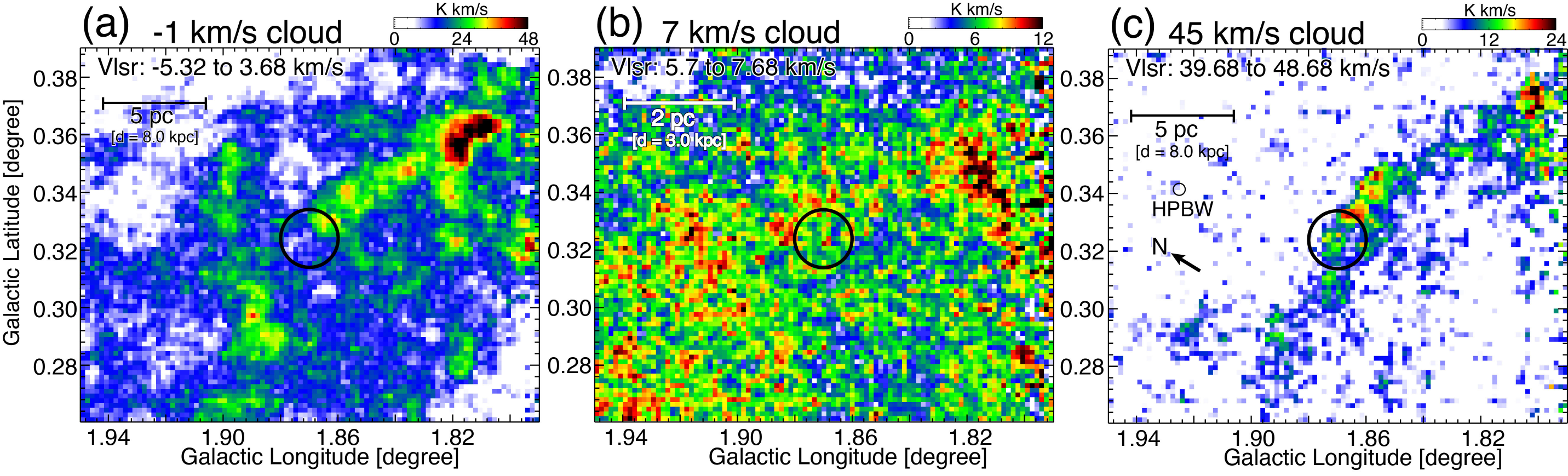}
\end{center}
\caption{Panels (a--c): Integrated intensity distributions of the three clouds in $\twelvecohh$ toward G1.9. The black circle indicates the position of the SNR. The integrated velocity ranges are indicated at the top-left of each panel. The scale bar, HPBW, and the north direction are indicated in the top left corner.}\label{clouds}
\end{figure*}
%%%%%%%%%%%%%%%%%%%%%%%%%%%%%%%%%

Figure~\ref{spec}a shows averaged line spectra of $\twelvecohh$, $\thirteencoh$, and OH 1667\,MHz toward G1.9.
The averaged area, indicated by the black square at the first panel in figure~\ref{lbch} corresponds with the grid size of the coarsest data (OH 1667\,MHz).
The line profile of $\twelvecohh$ emission peaks at $\sim$$-$1, 7, and 45~$\kms$.
Although the $\sim$$-$1~$\kms$ component appears to be a combination of two narrower sub-components at $-$1 and 3~$\kms$, the temperature decreases at 0~$\kms$ is less than three $\sigma$ and these two sub-components are not visible in $\thirteencoh$.
Thus, we consider the $-$1~$\kms$ component to be a single component.
The $-$1 and 7~$\kms$ components are inter-mingled in both, $\twelvecohh$ and $\thirteencoh$.
In $\thirteencoh$, the 7~$\kms$ component is strongest and clearly seen, while the 45~$\kms$ component is buried in noise and invisible.

Although the $-$1 and 7~$\kms$ components are mingled, it is more distinguishable in the $l$--$v$ diagram in figure~\ref{spec}b.
A cloud at 45~$\kms$ is isolated from the other clouds.
We hereafter call these three components $-$1, 7, and 45~$\kms$ clouds.

In figure~\ref{lbch} we show the velocity channel distribution on the large-scale view (0$\fdg$15 $\times$ 0$\fdg$12) toward the SNR.
The three molecular clouds exhibit individual and different morphological structures.
The $-$1~$\kms$ cloud is visible from $\vlsr$ = $-$16~$\kms$ and ends at $\vlsr$ = 9~$\kms$ with the filamentary shape extending $\sim$0$\fdg$10 from the north-east to the south-west.
This filament has an intensity depression in the direction of the SNR (see panels at $\vlsr$ from $-$11 to $-$1~$\kms$).
This cloud has the highest column density among the three molecular clouds.
The 7~$\kms$ cloud consists of diffuse weak emissions spreading over the entire field with a narrow velocity span (see blue-colored emissions at $\vlsr$ from 4.0 to 14.0~$\kms$).
While the integrated intensity of the cloud is the lowest among the three, the emitting area is the largest, and hence the averaged brightness temperature in the spectrum is very strong (see figure~2).
The 45~$\kms$ cloud is elongated and its filaments extend 0$\fdg$05 from east to west.
This cloud has a large velocity span comparable to the $-$1~$\kms$ cloud and these two clouds are morphologically alike.
However, the following differences suggest that the two clouds are independent features and not to be a unified cloud: (1) the slope angles of these two filaments are slightly different, (2) the integrated intensity of the $-$1~$\kms$ cloud is twice as high as that of the 45~$\kms$ cloud, and (3) these clouds are not connected in the velocity space (see the panel 29.0 to 34.0~$\kms$ in figure~\ref{lbch}), lending support to a chance overlapping of the two individual clouds.

In order to highlight the characteristics of these three clouds, we defined a velocity range, which represents the clouds' major morphological feature (hereafter the representative velocity range; $\vrep$).
\citet{eno21} developed the derivation of the representative velocity range of a cloud and showed that $\vcen$$\pm$$\dv$, where $\vcen$ and $\dv$ are the means of moment 1 and moment 2 values, respectively, can be a good expression of the $\vrep$ even for the case where two clouds are mingled with each other (the moments method).
By using the moments method, the representative velocity ranges for the $-$1, 7, and 45~$\kms$ clouds were derived to be $-$5.32 to 3.68, 5.7 to 7.68, and 39.68 to 48.68~$\kms$, respectively.
Figure~\ref{clouds} shows $\twelvecohh$ distributions of the three clouds obtained by integrating each $\vrep$.
Note that the maximum values of the color bars in the figure are at most four times different. 
The figure clearly exhibits the morphological characteristics of the three clouds.
Contrary to the line spectra, the 7~$\kms$ cloud has the weakest integrated intensity (i.e., column density) due to its very narrow velocity span.
%On the other hand, the $l$--$v$ diagram toward G1.9 in Figure~\ref{clouds}d exhibits the three clouds' velocity structures.
%In the Figure, the narrow line width of the 7~$\kms$ cloud also seen in Figure~2 is outstanding. 

\subsection{Distances to the molecular clouds}%3.2

%%%%%%%%%%%%%%%%%%%%%%%%%%%%%%%%%%%%%%%%%
\begin{figure*}[htbp]%Fig5	distance
\begin{center}
\includegraphics[width=15cm]{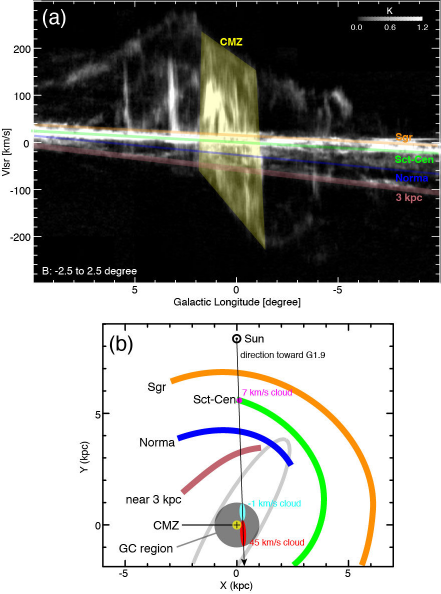}
\end{center}
\caption{(a) Longitude-velocity diagram of $\twelvecol$ toward the GC region, integrating latitude over $-$2$\fdg$5 to 2$\fdg$5. The CMZ and the known Galactic arms, based on \citet{rei16}, are drawn by the yellow and the other colors, respectively. (b) Schematic top-down view of the Galaxy based on \citet{rei16}. The black cross and filled, light gray circle indicate the GC, and the GC region, respectively. The CMZ and known Galactic arms are colored in the same manner as in panel (a). The possible location ranges of the $-$1, 7, and 45~$\kms$ clouds are indicated by cyan, magenta and red colors, respectively.}\label{dist}
\end{figure*}
%%%%%%%%%%%%%%%%%%%%%%%%%%%%%%%%%%%%%%%%%

The foreground clouds and clouds in the GC region (hereafter GC clouds) have different characteristics in space and velocity in $^{12}$CO. Thus, it is possible to estimate the line-of-sight (LOS) location of a cloud toward the GC region (see, e.g., \cite{eno14}) based on:
\begin{enumerate}[(i)]
	\item Following the Galactic rotation, the foreground clouds can be observable only in $-$60 $\leq$ $\vlsr$ $\leq$ 30~$\kms$, while velocities of the GC clouds have no such limitation due to non-circular motions (e.g., \cite{rei16}).
	\item The foreground clouds have typical line-widths of $\lesssim$ 5~$\kms$ in FWHM, while those of the GC clouds are much larger, originating from highly turbulent gas motions in this region and typically exceeding 10~$\kms$ (e.g., \cite{mor96}).
	\item $^{12}$CO is optically thick and cannot trace small, detailed structures within a molecular clump and filament. Given that the sizes of the clumps and filaments are a few pc \citep{gol87}, the foreground cloud is diffusively extended on 0$\fdg$1 scale because of its close distance, while the GC cloud consists of clumps and filaments on the same scale (e.g., \cite{eno14}).
\end{enumerate}

Although (i)--(iii) are not the absolute indicators but trends, (i) and (ii) in particular are visible in the large-scale $l$--$v$ diagram shown in figure~\ref{dist}a which includes clouds at all LOS distances toward the GC region.
The narrow velocity-width clouds indicated by transparent, orange, light-green, blue, and pink belts are the foreground clouds, whereas the others are the GC clouds (i.e., clouds within the Galactocentric radius of $\sim$1~kpc).
The marked cloud complex showing parallelogram shape, colored transparent, yellow parallelogram is a well-known dense gas complex within the Galactocentric radius of $\sim$200 pc, which is often referred to as the central molecular zone (CMZ; \cite{mor96}).
Based on this, G1.9 is located toward the eastern edge of the CMZ.

In accordance with the assessment using (i)--(iii), it is most likely that the 7~$\kms$ cloud is located in the foreground.
From its velocity, this cloud is likely located in Scutum-Centaurus (Sct-Cen) Arm and thus the distance is $\sim$3~kpc (\cite{rei16,vel12}; see figures~\ref{dist} and \ref{spec}b).

The velocity of the $-$1~$\kms$ cloud is between Norma and Scutum arms, and it has quite a large velocity width exceeding 20~$\kms$.
Although the large line width supports the location of the GC region, its velocity is close to that of the foreground clouds. 
Molecular clouds with such a large velocity width are only observed in the Galactic superbubbles or $\htwo$ regions except for the GC clouds (e.g., \cite{fuk99}).
Thus, we searched catalogues and an excess of infrared emission toward the G1.9 region, but no superbubbles and $\htwo$ regions have been detected within the 0$\fdg$15 $\times$ 0$\fdg$12 field shown in figure~\ref{lbch}.
Furthermore, there are no known foreground clouds with the velocity of $-$1~$\kms$ (see figure~\ref{dist}a), and the morphology of the $-$1~$\kms$ cloud is filamentary.
Therefore, we suggest its location in the GC region.

The 45~$\kms$ cloud is most likely located at the distance of the GC region because it has a large velocity width ($\sim$15~$\kms$ in FWHM), the velocity is far away from that of the foreground clouds, and the morphology is filamentary.

Since G1.9 is located toward the edge of the CMZ, and the boundary of the CMZ is ambiguous, possible locations of the two GC clouds (i.e., $-$1 and 45~$\kms$ clouds) are on the near side of the GC region, in the CMZ, and on the far side of the GC region.
\citet{saw04} suggested a method to estimate the LOS distance for a GC cloud based on CO and OH emission/absorption.
According to \citet{saw04}, the CMZ ($|l| \leq 2\fdg0$) is a bright 18-cm continuum source, and hence the 1667~MHz OH line shows absorption if the cloud is located in the nearside of the GC region.
Since the angular resolution of our 18 cm data is poor, and we cannot detect significant emission from G1.9, the background source for the absorption in our data is likely to be the CMZ.

We found deep absorption from $-$60 to $+$40~$\kms$ in our OH data (see figure~\ref{spec}a).
The spectrum is noisy, but its smoothed spectrum shows the deepest absorption at $\sim$6~$\kms$, and it also shows a tail in the negative velocity, thus this absorption is likely caused by the $-$1 and 7~$\kms$ clouds.
Thus, the $-$1~$\kms$ cloud is possibly located at the near side in the GC region (see figure~\ref{dist}b).
Since the $-$1 and 45~$\kms$ clouds are $not$ a unified feature, the 45~$\kms$ cloud, which does not show significant absorption in OH, is probably located in the bright background OH source (i.e., the CMZ) or in the far side of the GC region (see figure~\ref{dist}b).
Assuming the distance to the GC is 8.3~kpc (e.g., \cite{gil09}), the distances $d$ to the $-$1, and 45~$\kms$ clouds are 7 $\lesssim$ $d$ $\le$ 8.1~kpc and 8.1 $\le$ $d$ $\lesssim$ 9~kpc, respectively (see figure~\ref{dist}b).

Given that the $-$1~$\kms$ cloud is located at its projected distance ($i. e.,$ $\sim$270~pc) close to us from the GC, we hereafter use 8.0~kpc as its distance.
The possible positions and LOS-distance ranges of the three clouds in the top-down view of the Galaxy are illustrated in figure~\ref{dist}b.

\subsection{Physical parameters of the clouds}	%3.3

%%%%%%%%%%%%%%%%%%%%%%%%%%%%%%%%%%%%%%%%%%%%%%%%%%%%%%%%%%

\begin{table*}[!htbp]	%table2
\tbl{Physical parameters of the molecular clouds toward G1.9}{%
\begin{tabular}{lccccc}
\hline\noalign{\vskip3pt}
      Name & $\vrep$ & $\vlsr$ & distance & \nhtwo ~(typical/peak) & Mass \\
       & ($\kms$) & ($\kms$) & (kpc) & ($\times$10$^{21}$ cm$^{-2}$) & ($\times$10$^{3}$ \msun)\\
\hline\noalign{\vskip3pt} 
$-$1~$\kms$ cloud &$-$5.32 to 3.68 &  $-$18.3 to 5.7 & 7 $\lesssim$ $d$ $\le$ 8.1 & 3.3\footnotemark[$*$] / 9.4\footnotemark[$*$] & 4.6\footnotemark[$*$]  \\
7~$\kms$ cloud &  5.7 to 7.68 &  5.7 to 9.7 &$\sim$3 & 2.4 / 5.5 & 3.8  \\
45~$\kms$ cloud &  39.68 to 48.68 &  34.7 to 55.7 & 8.1 $\le$ $d$ $\lesssim$ 9 & 0.5 / 4.2 & 1.0  \\
$-$1~$\kms$ cloud within the shell &  &$-$18.3 to 5.7 & 7 $\lesssim$ $d$ $\le$ 8.1 & $\sim$3 / $\sim$6 & 0.6  \\
\hline\noalign{\vskip3pt} 
\end{tabular}}
\label{tab:phys}
\begin{tabnote}
\hangindent6pt\noindent
\footnotemark[$*$] lower limits. Note. --- Col.1, names of clouds: Col.2, representative velocity ranges for the clouds: Col.3, velocity ranges used for the column density and mass derivations: Col.4, distances to the clouds: Col.5, typical and peak molecular column densities: Col.6, molecular masses.\\
\end{tabnote}
\end{table*}

%%%%%%%%%%%%%%%%%%%%%%%%%%%%%%%%%%%%%%%%%%%%%%%%%%%%%%%%%%

In order to estimate a column density and mass, we summarized $\vlsr$ and $\vrep$ of clouds in table~\ref{tab:phys}.
Note that $\vrep$ is a limited velocity range, symbolizing the representative shape of a cloud, and does not include all the emission that makes up the cloud.
To derive the column density and mass, ideally we need to use the end-to-end, full velocity range (\vlsr), which includes all the emission, and the emission in that velocity range need to be dominated by only one cloud.
From the detailed velocity channel distribution and the $l$--$v$ diagram in figure~\ref{spec}b, we estimated these full velocity ranges for the $-$1, 7, and 45~$\kms$ clouds to be $-$18.3 to 11.7, 5.7 to 9.7, and 34.7 to 55.7~$\kms$, respectively.
In 5.7 to 9.7~$\kms$, the $-$1 and 7~$\kms$ clouds are mingled with each other.
Therefore, we determined the end of the velocity range for deriving the column density and mass for the $-$1~$\kms$ cloud to be 5.7~$\kms$, where the 7~$\kms$ cloud begins to dominate.
Thus, the derived column density and mass for the $-$1~$\kms$ cloud give a lower limit.

For the $-$1 and 45~$\kms$ clouds, we use the typical value of $\twelvecohh$/$\twelvecol$ as 0.7 \citep{oka12} and of the conversion factor ($X_\mathrm{CO}$) as 0.7 $\times$10$^{20}$ cm$^{-2}$ (K~km~s$^{-1}$)$^{-1}$ \citep{tor10}, while for the 7~$\kms$ cloud we use $\twelvecohh$/$\twelvecol$ as 0.4 \citep{oka12} and $X_\mathrm{CO}$ as 1.0 $\times$10$^{20}$ cm$^{-2}$ (K~km~s$^{-1}$)$^{-1}$ \citep{oka17}.
A column density of molecular hydrogen is given by equation (1),

\begin{equation}
  \nhtwo = X_\mathrm{CO} \times W(CO),
\end{equation}
where $W$(CO) is the integrated intensity of $\twelvecol$.
To estimate the hydrogen mass, we use the following equation:

\begin{equation}
  M = \mu m_p \sum_{i} ~[d^2 \Omega \nhtwo_{,i}],
\end{equation}
where $\mu$, $m_p$, $d$, $\Omega$ and $\nhtwo_{,i}$ are the mean molecular weight, proton mass, distance, a solid angle subtended by a pixel, and column density of molecular hydrogen for the $i$-th pixel, respectively.
We assume a helium abundance of 20$\%$, which corresponds to $\mu$=2.8.
The deduced peak and typical column densities, and masses of the clouds are summarized in table~\ref{tab:phys}.

The areas used for the mass estimation for the $-$1, 7, and 45~$\kms$ clouds are the 0$\fdg$1 filament, whole the field, and the 0$\fdg$05 filament in figure~\ref{clouds}.
Thus, given the 0$\fdg$1 filament extends more from the lower Galactic latitude (i.e., $<$ 0\fdg29), the mass of the $-$1~$\kms$ cloud is the lower limit.
The physical parameters for the $-1$~$\kms$ cloud within the shell in table~\ref{tab:phys} were estimated in the area enclosed by the inner and outer boundaries of the shell (see magenta circles in figure~\ref{g1.9}).

%============================Results (above)=======================

%============================Discussion (below)=======================
\section{Discussion}
\subsection{Distance to G1.9 and its associated cloud(s)}%4.1
We first focus on the morphological coincidences between the three clouds with the SNR.
Among the three clouds, the 7~$\kms$ cloud does not show any morphological correlation with the SNR, whereas the $-$1~$\kms$ cloud shows the obvious gas depression toward the SNR G1.9 (figures~\ref{lbch}--\ref{clouds}).
This might be evidence of the interaction between the SNR and the $-$1~$\kms$ cloud.
The 45~$\kms$ cloud has a filamentary shape and is overlapping the SNR, thus this cloud also is possibly associated with the SNR.

Next, we discuss the distances as the indicator of the association.
The estimated distances to the $-$1, 7 and 45~$\kms$ clouds are 7 $\lesssim$ $d$ $\le$ 8.1~kpc, $\sim$3~kpc, and 8.1 $\le$ $d$ $\lesssim$ 9~kpc, respectively (see also Figure~\ref{dist}b).
On the other hand, the distance to G1.9 is bit uncertain.

\citet{gre84} discovered G1.9 and first suggested its distance to be $<$20~kpc.
\citet{nor04} carried out wide-field imaging of the GC at 330~MHz, and by comparing 74~MHz flux they found that G1.9 does not have significant 74~MHz absorption.
Thus, to avoid absorption, they concluded that the SNR may be located on the near side of the GC region, i.e., $<$7.8~kpc \citep{nor04}.
\citet{rey08} measured an extremely high absorbing column density of 5.5 $\times$ 10$^{22}$ cm$^{-2}$ toward G1.9, and thus suggested that the location is at least $not$ in the foreground (hence in the GC region or farther).
Based on a stellar synthesis model, half of the extinction is accounted for by material located in front of the GC region \citep{rey08}.
On the other hand, if the distance is far from the GC, the derived expansion velocity is unnaturally high.
Thus, \citet{rey08} finally concluded that the location in the GC region is the most plausible.
Note that this distance requires high column density accounting for the remaining half of the extinction to the associated cloud, even though they could not find it.
\citet{bor10} measured line widths and confirmed that the values are consistent with the expanding velocity if assuming the distance of 8.5~kpc (14,000~$\kms$).
Furthermore, \citet{car11} monitored the X-ray filaments and confirmed that the shock velocity assuming 8.5~kpc is comparable to the spectroscopically deduced velocity.
There is another estimated distance by \citet{roy14} who presented HI absorption measurements and suggested that G1.9 is 2~kpc further away than the GC.
However, we did not find the absorption in the archival HI combined data obtained with ATCA and Parkes \citep{mcc12}.
In addition, \citet{roy14} does not include information regarding data or the estimation method.
Therefore, we do not consider the distance suggested by \citet{roy14} further.

Although the distance estimated by \citet{nor04} is inconsistent with the estimated distance to the 45~$\kms$ cloud, almost all of these arguments suggest that the $-$1 and 45~$\kms$ clouds are possible counterparts to the SNR.
However, taking into account that the $-$1 and 45~$\kms$ clouds are not associated with each other (see subsection 3.1), only one of these two clouds is possibly associated with the SNR.

Since the estimated distance to the $-$1~$\kms$ cloud agrees with all the estimated distances to G1.9 in the literature, and the cloud shows a clear depression of gas toward the SNR than 45~$\kms$ cloud (see figure~3), the $-$1~$\kms$ cloud is the more favorable counterpart.
Therefore, hereafter we call the $-$1~$\kms$ cloud the $associated$ $cloud$ with respect to the SNR and use 8.0~kpc as the distance to G1.9.
However, we still cannot completely rule out the possibility of a physical association between the 45~$\kms$ cloud and G1.9.
According to \citet{set04}, a broad velocity wing is the strongest evidence for the association, while we found no significant differences between spectra in the SNR region and the other region.
This may be due to beam dilution by small emitting areas of the wing cloud for this small, young SNR.
Further investigations such as a detection of the broad velocity wing, and comparison of distributions of the SNR and gas temperature or excitation state derived by future multi-line observations are required to obtain robust evidence.

\subsection{Spatial-velocity distribution of the associated cloud}%4.2

%%%%%%%%%%%%%%%%%%%%%%%%%%%%%%%%%%%%%%%%%%%%%%%%%%%%%%%%%%%%%%
\begin{figure*}[t]%Fig6	LBch
\begin{center}
\includegraphics[width=15cm]{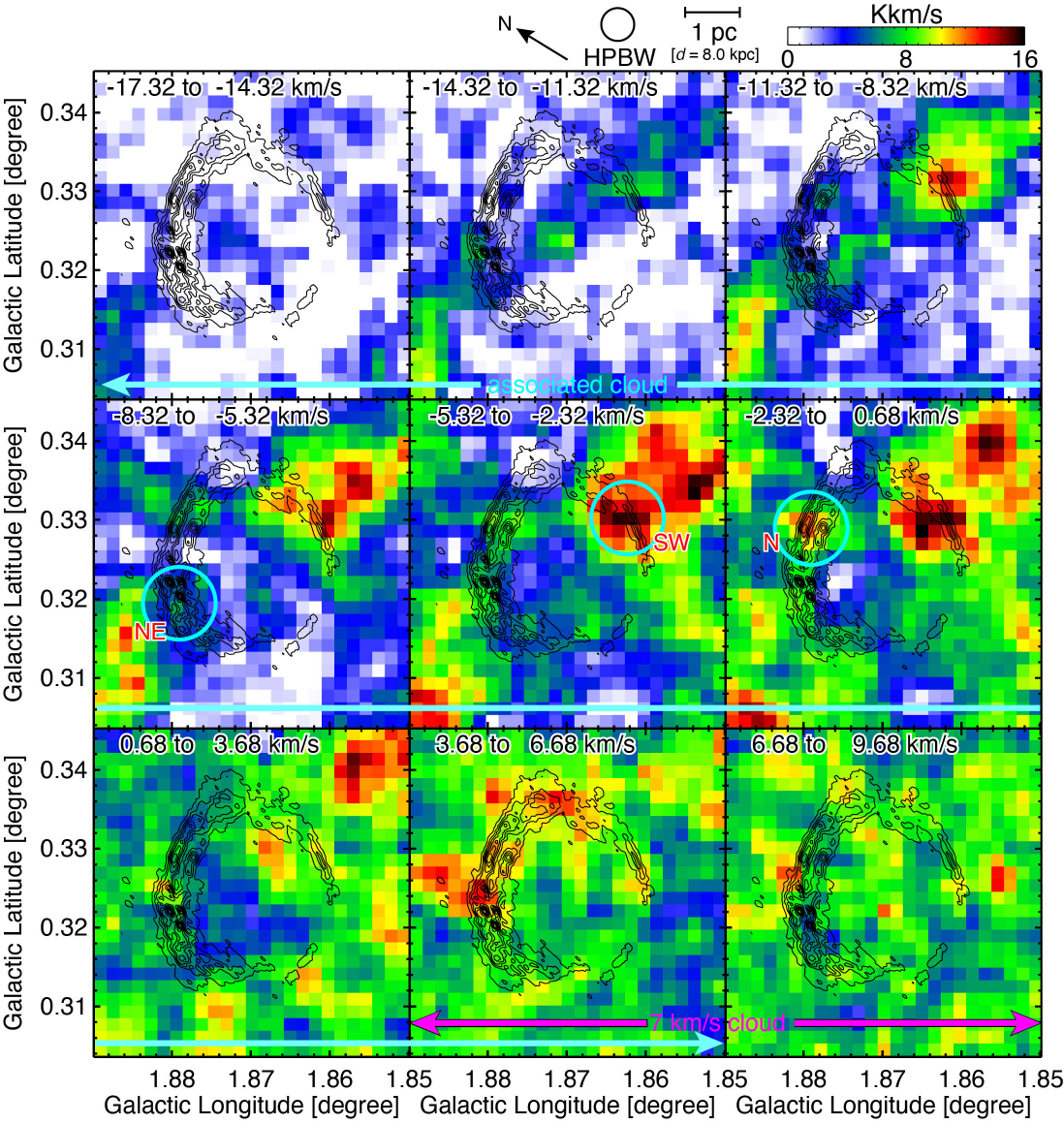}
\end{center}
\caption{Velocity channel distribution of $\twelvecohh$ toward the associated cloud obtained with JCMT, overlaid with the black contours outlining 9~GHz radio-continuum emission. The cyan circles indicate the positions of emission enhancements toward the shell of the SNR. The half-power beam-width (HPBW) and the direction of north are indicated at the top of the figure. The velocity ranges for the clouds are indicated by cyan and magenta arrows at the bottom of the panels.}\label{albch}
\end{figure*}
%%%%%%%%%%%%%%%%%%%%%%%%%%%%%%%%%%%%%%%%%%%%%%%%%%%%%%%%%%%%%%

In figure~\ref{albch} we show the velocity channel distribution of the associated cloud in $\twelvecohh$ overlaid with black contours of the 9~GHz emission.
The filamentary structure of the associated cloud is clearly seen from $\vlsr$ = $-$14.32 to $-$8.32~$\kms$.
At the velocity from $-$8.32 to 0.68~$\kms$, the emissions inside the SNR are depressed and the filament is discontinued there.
The south-western (SW), northern (N), and north-eastern (NE) radio-bright rims coincide well with local enhancements of the molecular gas emission at $-$5.32 to $-$2.32~$\kms$, $-$2.32 to 0.68~$\kms$, and $-$8.32 to $-$5.32~$\kms$, respectively (see cyan circles in figure~\ref{albch}), whereas molecular gas is barely detected toward the south-eastern direction of the SNR where the radio continuum emission is faint.

As the maximum energy is $E_\mathrm{max} \propto B_{\nu} v^{2}_\mathrm{shock}$, if dense molecular gas surrounding the SNR interrupts the expansion of the SN shock, such a region would be brighter in radio rather than X-ray \citep{bor17}.
%Although we cannot lead strong conclusion because 
While the current angular resolution of CO is quite coarse compared to radio and X-ray, the coincidences of local peaks between CO and radio may suggest the interaction between the gas and the SNR.
Through point-by-point comparisons with higher angular resolution CO data and the other wavelengths, the validity of the interaction can be examined.
At the velocity range from 3.68 to 9.86~$\kms$, it is difficult to find the associated cloud because of the overlapping of the foreground diffuse cloud in the Sct--Cen~Arm (i.e., the 7~$\kms$ cloud).

%%%%%%%%%%%%%%%%%%%%%%%%%%%%%%%%%%%%%%%%%%%%%%%%%%%%%%%%%%%%
\begin{figure*}[t]%Fig7	LBch
\begin{center}
\includegraphics[width=15cm]{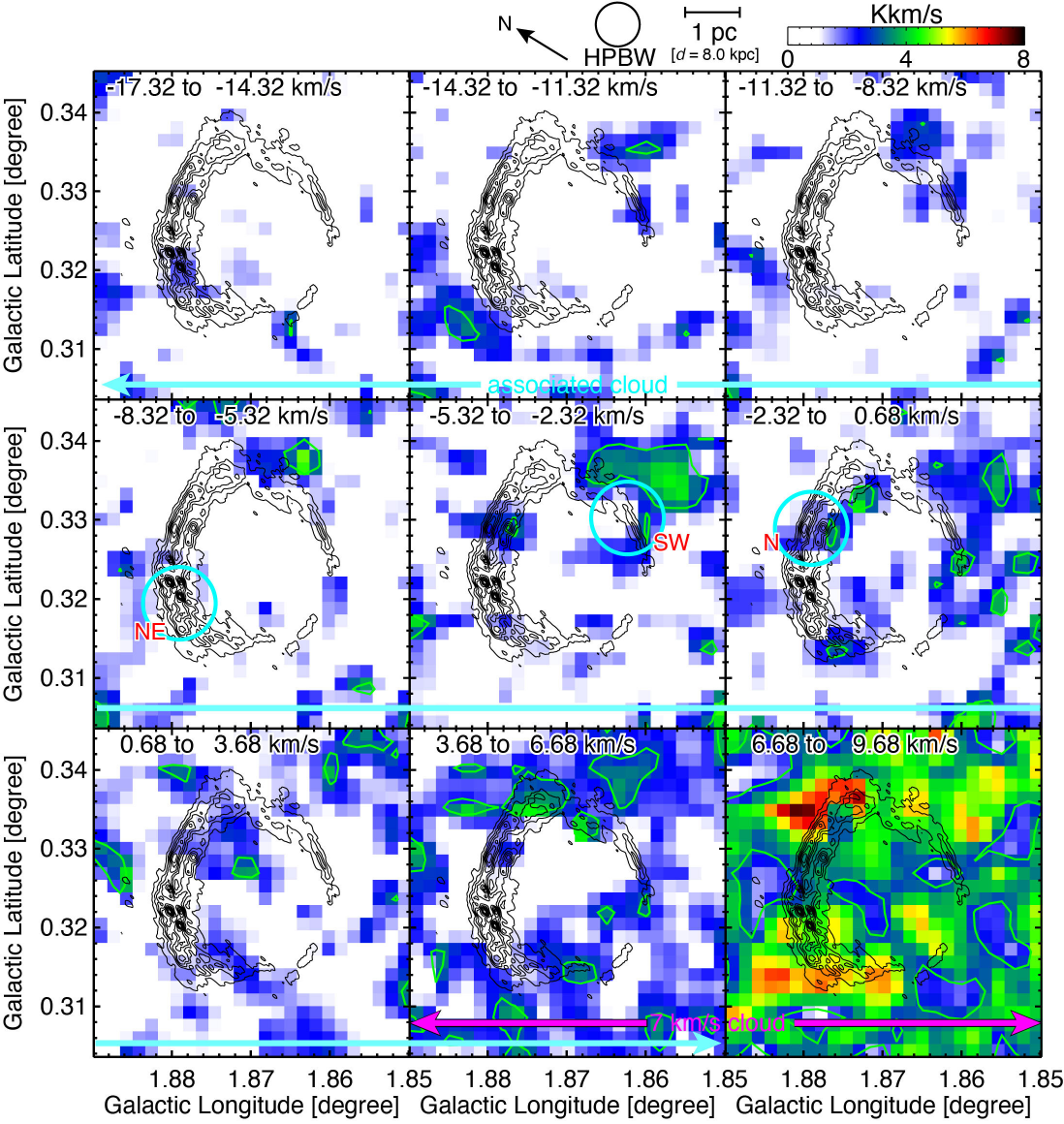}
\end{center}
\caption{Velocity channel distribution of $\thirteencoh$ toward the associated cloud obtained with APEX, overlaid with black and light green contours outlining 9~GHz radio-continuum emission and a three sigma level $\thirteencoh$ emission (=2.87 Kkm/s), respectively. The cyan circles indicate the positions of enhancements in $\twelvecohh$ toward the shell of the SNR (see figure~\ref{albch}). The HPBW and north are indicated at the top of figure. The velocity ranges for the clouds are indicated by cyan and magenta arrows at the bottom of the panels.}\label{13co}
\end{figure*}
%%%%%%%%%%%%%%%%%%%%%%%%%%%%%%%%%%%%%%%%%%%%%%%%%%%%%%%%%%%%

We next examine denser gas distribution toward the same region in figure~\ref{albch} by using $\thirteencoh$ data obtained with the SEDIGISM survey \citep{sch21}. As shown in light green contours in figure~\ref{13co}, which correspond a 3$\sigma$ significance level, compared to $\twelvecohh$, $\thirteencoh$ emission is very weak in the associated cloud, whereas that from the 7~$\kms$ cloud is very strong.
According to \citet{tok19}, the intensity ratio of $\riso$ is significantly higher in the foreground clouds than the GC clouds (see also figure~10 in the Appendix), which further supports the associated cloud's location in the GC region.

In the associated cloud, there is a tendency that the bright $\thirteencoh$ areas, which correspond to denser gas, are located outside the SN shell, whereas $\twelvecohh$ emissions corresponding to high temperature and/or dense gas are distributed both at the shell and outside.
This may indicate that the molecular gas in the shell is heated by the shock, while the gas outside the shell has not yet experienced an interaction with the SNR and hence maintains the high density.

%%%%%%%%%%%%%%%%%%%%%%%%%%%%%%%%%%%%%%%%%%%%%%%%%%%%%%%%%%%%
\begin{figure*}[t]%Fig8	XYV diagram
\begin{center}
\includegraphics[width=15cm]{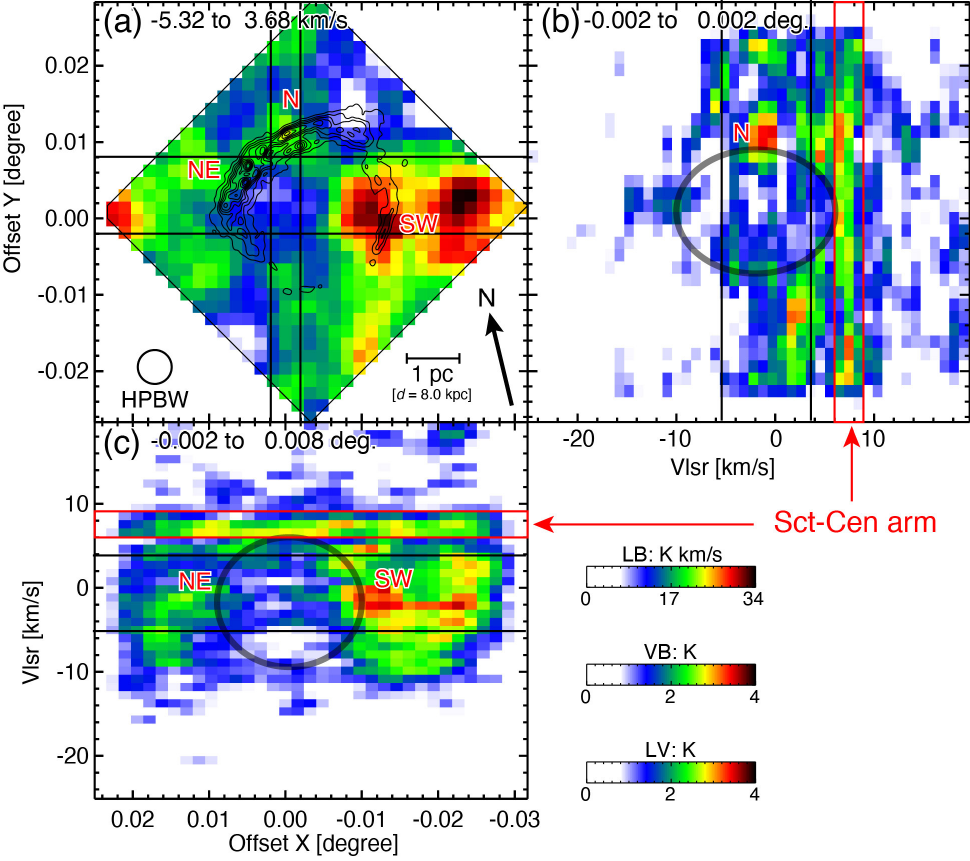}
\end{center}
\caption{(a) Integrated intensity distribution of the associated cloud in $\twelvecohh$ overlaid with the contours of 9~GHz continuum emission in the Offset X--Offset Y coordinates. The direction of north, the HPBW, and the scale bar are indicated at the bottom corners of the panel. The integration ranges for panels (b) and (c) are indicated by the black lines. (b) Velocity--Offset Y diagram of the associated cloud. The integrated velocity ranges for panel (a) and the Sct-Cen arm are indicated by the white and red lines, respectively. (c) Offset X--Velocity diagram of the associated cloud. The integrated velocity ranges for panel (a) and the Sct-Cen arm are indicated by the white and red lines, respectively.}\label{xyv}
\end{figure*}
%%%%%%%%%%%%%%%%%%%%%%%%%%%%%%%%%%%%%%%%%%%%%%%%%%%%%%%%%%%%

In figure~\ref{xyv} we show integrated intensity distribution and position--velocity diagrams of the associated cloud in $\twelvecohh$ in the Offset~X--Offset~Y~coordinates, which is defined as rotated 45 degrees clockwise from the Galactic coordinates centered at ($l$, $b$) = (1\fdg870, 0\fdg325).
Most emission from the associated cloud coincident with the SN shell is seen in the N, NE, and SW directions, where the radio-bright rims are found [see panels~(a)--(c)].
We also found a cavity-like structure in each position-velocity diagram of CO whose velocity range is from $-$10 to 6~$\kms$ [see black ellipses in panels (b) and (c)].
Note that the spatial extent of the CO cavity is roughly consistent with that of the radio continuum shell.
This is further possible evidence for the interaction between the cloud and the SNR, because such a cavity-like structure, corresponding to an expanding gas motion, is thought to be formed by supernova shocks and/or strong winds from the progenitor system of the SNR (e.g., \cite{koo90,koo91}).

We argue that the gas acceleration due to supernova shocks is negligible in G1.9.
As described in subsection~3.3, the mass of the $-$1~$\kms$ cloud within the shell is $\sim$600$\msun$.
If the ambient gas with this large mass was uniformly filled over the current volume of the SNR before being blown out, the initial ambient density is estimated to be $\sim$660\,cm$^{-3}$.
By contrast, previous X-ray spectroscopic studies indicated the low pre-shock density of $\sim$0.04 cm$^{-3}$ \citep{rey08}, based on the high velocity of the SNR forward shock \citep{bro19} and low ionization state of the post-shock gas.
This discrepancy implies that the expanding CO gas was first formed by the strong pre-explosion winds and subsequently the progenitor of G1.9 exploded inside the low-density cavity.
The smaller wind bubble compared to that seen in the disk is expected to be due to the high gas density in the GC.
Since G1.9 is widely thought to be a Type~Ia supernova remnant (e.g., \cite{rey08,bor10,bor13,cha16,bor17,luk20,gri21}), such strong pre-explosion winds can be only seen in a progenitor system comprising a white dwarf and non-degenerate companion star (also known as a single-degenerate system).

An alternative idea is that the expanding gas was formed by the strong stellar wind from the high-mass progenitor of G1.9.
According to \citet{luk20}, the observed rotation measure can be explained as a combination of the red supergiant wind and toroidal magnetic field hypothesis.
In case G1.9 is a core-collapse remnant, the expansion velocity of CO gas $\sim$8~$\kms$ is roughly consistent with the other core-collapse SNRs with stellar wind bubbles (e.g., \cite{fuk12,kur18,san21}).
In any case, this further supports the association between the $-$1~$\kms$ cloud and the SNR.

\subsection{Comparison with the X-ray and radio continuum}%4.3

%%%%%%%%%%%%%%%%%%%%%%%%%%%%%%%%%%%%%%%%%%%%%%%%%%%%%%%%%%%%%%%
\begin{figure*}[t]%Fig9	Az profile
\begin{center}
    \includegraphics[width=15cm]{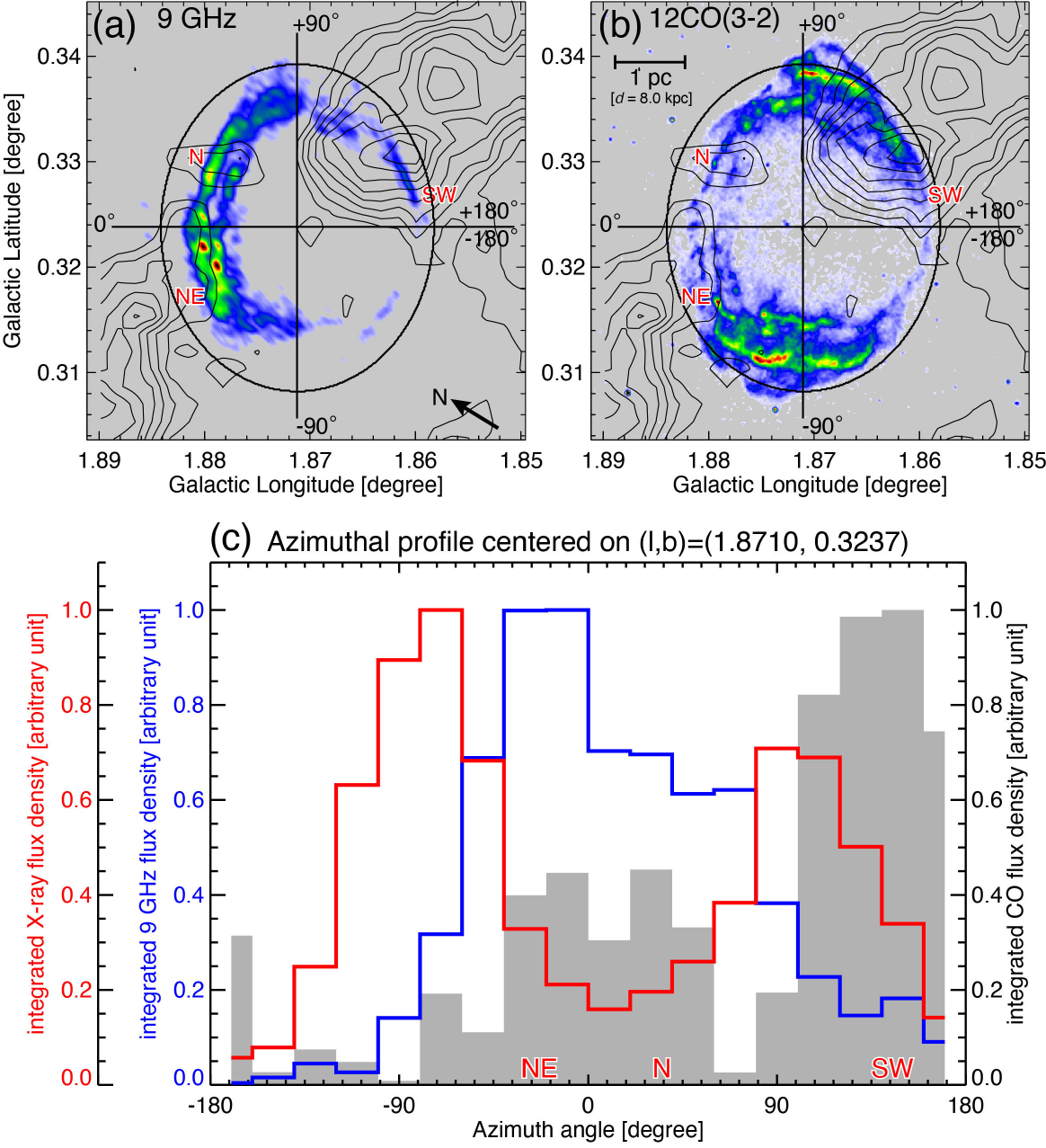}
  \end{center}
  \caption{(a) Distribution of 9~GHz continuum emission with overlaid contours of the integrated intensity distribution of the associated cloud in $\twelvecohh$. The integration range corresponds to $\vlsr$ in table 2 ($-$18.3 to 5.7~$\kms$). The interval and the lowest level of the contours are 4 and 30 $\kkms$, respectively. The black ellipse indicates the area calculating the azimuthal profile shown in panel (c). (b) Distribution of X-ray emission with the same CO contours in panel (a). (c) Azimuthal profiles compiled flux inside the black ellipse indicated in panel (a) centered at the dynamical center of the SNR derived by \citet{bor17} with the radius and eccentricity of 0$\fdg$013 and 0.55, respectively. The filled grey area and the areas marked by blue and red lines indicate $\twelvecohh$, 9~GHz continuum, and X-ray emissions normalized to the peak, respectively. Pixels (voxels) below 5$\sigma$ were removed from each data in advance.}\label{az}
\end{figure*}
%%%%%%%%%%%%%%%%%%%%%%%%%%%%%%%%%%%%%%%%%%%%%%%%%%%%%%%%%%%%%%%

%In order to discuss the origin of radio and X-ray emissions, we here compare spatial distributions of radio, X-ray and CO in more detail.
In figures \ref{az}a and \ref{az}b we show distributions of 9~GHz continuum and X-ray, respectively with the black contours showing $\twelvecohh$.
We found that the three intensity peaks of the radio-bright rims as NE, N, and SW, appear to be overlapped with molecular clouds as shown by black contours, while the radio-dim regions have no strong CO emission.
However, synchrotron X-rays are bright in the CO-dim regions except for the SW rim.
Figure \ref{az}c shows that the spatial relations among the radio continuum, X-ray and CO are clearer than each overlay map.

We argue that this tendency could be explained by shock interactions with inhomogeneous gas distributions, which was first proposed by \citet{bor17}.
The ambient gas density in the northern radio-brightest rim is at least 10 times larger than that of the eastern X-ray-brightest rim (see figure~\ref{az}c).
Since the shock velocity $v_\mathrm{shock}$ is inversely proportional to the square root of gas density, the shock velocity of the eastern rim is expected to be at least three times higher than that of the northern rim.
Indeed, the previous measurements of the X-ray proper motion indicated that the X-ray-bright major axis (east-west) is $\sim$3--4 times faster than the northern radio bright shell \citep{bor17}.
Since the maximum energy of accelerated cosmic-ray electrons $E_\mathrm{max} \propto B_{\nu} v^{2}_\mathrm{shock}$ under the age-limited acceleration (e.g., \cite{rey08}), the shock-velocity difference due to the inhomogeneous gas distribution is naturally expected. 

The shock interactions with the clumpy medium can explain the origin of the SW X-rays-bright rim despite a large amount of gas.
According to magnetohydrodynamic (MHD) numerical simulations, shock-cloud interactions can enhance turbulent magnetic field up to $\sim$1~mG on the surface of the shocked clouds (e.g., \cite{ino12,cel19,pav18}).
The magnetic field amplification induces the synchrotron X-ray limb-brightening around the shocked clumps (e.g., \cite{san10,san13,tan20}).
In the case of G1.9, the SW CO clouds are expected to be highly clumpy with a clump size of $\sim$0.1~pc or less (e.g., \cite{san20}).
Further ALMA observations with a high-angular resolution and unprecedented sensitivity are needed to better understand the X-ray and radio continuum emission as well as the cosmic-ray electron acceleration to higher energy in the SNR G1.9 \citep{san15}.

\subsection{Implications for future observations}%4.4
Our analyses have revealed molecular gas possibly interacting with the SN shock toward the radio-bright rims (N, NE, SW).
According to \citet{aha17}, the increasing number of detections of so-called $\pi$$^{0}$-decay bump in the GeV/TeV spectra in SNRs has been reported, and such SNRs have correlations between distributions of TeV and gas emissions (e.g., \cite{fuk12}).
This has been considered to be substantial evidence for the acceleration of cosmic-ray protons in SNRs.
The molecular gas discovered in the present work may be responsible for the targets for hadronic interactions in G1.9, if hadronic-origin $\gamma$-rays are detected in the future, and hence investigation of the gas is essential.
Therefore, the comparison of the interacting gas candidates with the future gamma-ray data with high sensitivity and high angular resolution is an important perspective.

High--resolution, multi-line observations of the candidate clouds will allow us to investigate very accurately the interaction of gas with the high-velocity X-ray filaments expanding over 10,000~$\kms$.
This will reveal the effect of the SN-shock deceleration quantitatively.
Thanks to its very high SN shock velocity, G1.9 is the best and only laboratory to test physics of the SN shock propagation.
Therefore, a few to tens of years monitoring observations of the SNR in radio, X-ray and CO could provide new insights into gas dynamics and shock propagation.
%============================Discussion (above)=======================

%============================Conclusions (below)==============================
\section{Conclusions} \label{sec:conc}
We investigated the interstellar gas toward the youngest known Galactic SNR G1.9$+$0.3 by mainly using archival $\twelvecohh$ data obtained with the CHIMPS2 survey.
Based on the very large velocity width, exceeding 20~$\kms$ in $\twelvecohh$, we suggest the $-$1~$\kms$ cloud, whose estimated distance is 8.0~kpc, is possibly associated with the SNR.
The cloud has three peaks at N, NE, and SW areas of the SNR, which coincides well with the radio-bright rims, whereas the N and NE peaks are anti-correlated with the X-ray-bright rims.
The SW CO peak corresponding to the X-ray-bright rim can be interpreted as the result of the shock-cloud interaction.
The CO distribution is direct evidence that the anisotropic expansion of the observed SN shocks originates from the deceleration by interaction with the surrounding dense, anisotropic cloud.
%============================Conclusions (above)==============================

\section*{Funding}
This work was financially supported by Grants-in-Aid for Scientific Research (KAKENHI) from the Japanese Society for the Promotion for Science (JSPS; grant number society for 20K14520). 
MDF acknowledge Australian Research Council funding through grant DP200100784.

\section*{Conflict of Interest}
The authors declare that they have no conflict of interest.

\begin{ack}
We thank the anonymous referee(s) for helpful comments that improved the manuscript.
RE is grateful to Dr. K. Torii for providing the OH data.
The James Clerk Maxwell Telescope is operated by the East Asian Observatory on behalf of The National Astronomical Observatory of Japan; Academia Sinica Institute of Astronomy and Astrophysics; the Korea Astronomy and Space Science Institute; Center for Astronomical Mega-Science (as well as the National Key R\&D Program of China with No. 2017YFA0402700).
Additional funding support is provided by the Science and Technology Facilities Council of the United Kingdom and participating universities and organizations in the United Kingdom and Canada.
The scientific results reported in this article are based on data obtained from the Chandra Data Archive (Obs IDs: 6708, 8521, 10111, 10112, 10928, 10930, 12689, 12690, 12691, 12692, 12693, 12694, 12695, 13407, 13509, 16947, 16948, 16949, 17651, 17652, 17663, 17699, 17700, 17702, 17705, and 18354).
This research has made use of the software provided by the Chandra X-ray Center (CXC) in the application packages CIAO (v 4.12).
%This work was supported by JSPS KAKENHI Grant Numbers JP19H05075 (H. Sano), and JP21H01136 (H. Sano).
\end{ack}

\vspace{5mm}
%\software{CIAO \citep[v4.12:][]{fru06}, CALDB \citep[v4.9.1:][]{gra07}, miriad \citep{1995ASPC...77..433S}, karma \citep{1995ASPC...77..144G}}

%============================Appendix (below)==============================
\appendix
\section*{$\riso$ variations in the GC clouds and the foreground Galactic arms}

%%%%%%%%%%%%%%%%%%%%%%%%%%%%%%%%%%%%
\begin{figure*}[t]%Fig10	R1312
\begin{center}
    \includegraphics[width=15cm]{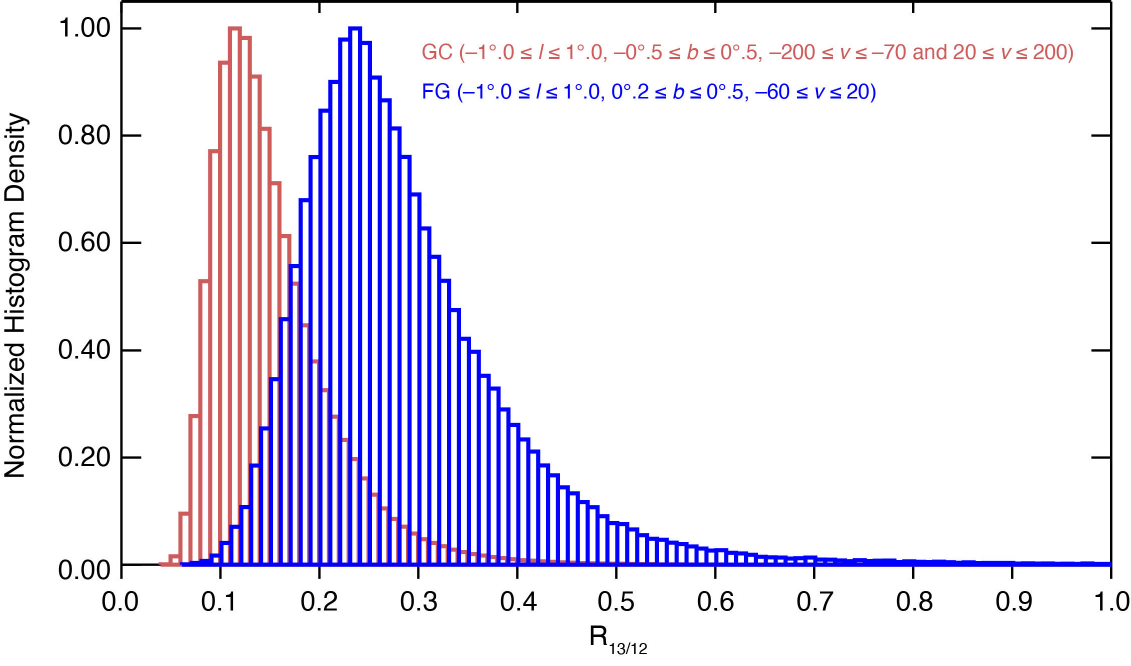}
  \end{center}
  \caption{Normalized histograms of $\riso$ for the GC gas (salmon pink) and foreground gas (blue). Voxels below 5$\sigma$ were removed from each data in advance.}\label{riso}
\end{figure*}
%%%%%%%%%%%%%%%%%%%%%%%%%%%%%%%%%%%%

Here, we present $\riso$ in the GC clouds and the other clouds in the foreground Galactic arms.
We used $\twelvecol$ and $\thirteencol$ data obtained simultaneously with the Nobeyama 45-m telescope by \citet{tok19}.
These data cubes have the same HPBWs, grid sizes, and coverages, thus they can be directly compared to each other.
We first flagged voxels having 5$\sigma$ and lower emissions from both data cubes to avoid noise fluctuations.

Next, we examined spectra in the two data cubes and determined two groups---one dominated by emission from the GC clouds to be ($l$, $b$, $v$) = ($-$1\fdg0 to 1\fdg0, $-$0\fdg5 to 0\fdg5, $-$200 to $-$70 and 20 to 200~$\kms$), and the other dominated by emission from the foreground clouds to be ($l$, $b$, $v$) = ($-$1\fdg0 to 1\fdg0, 0\fdg2 to 0\fdg5, $-$60 to 20~$\kms$).
The emission from the GC region is brighter than that from the foreground region, and thus the GC emission becomes absorption when the position and velocity of the two emissions match (i.e., $b$ = -0\fdg2 to 0\fdg2 and $v$ = $-$60 to 20~$\kms$).
This absorption is clearly visible in the $l$--$v$ diagram of $\twelvecol$.
Therefore, such voxels were excluded from the above $lbv$ definition.

Comparing these data cubes, we have obtained normalized histograms of $\riso$ for the foreground clouds (blue) and GC clouds (salmon pink) in figure~\ref{riso}.
As mentioned by \citet{tok19}, the foreground clouds have twice as much $\riso$ than the GC clouds, and hence these two location clouds are distinguishable by using $\riso$.

\end{document}